\documentclass{eas}

\usepackage{graphicx}
\usepackage{natbib}
\usepackage{amssymb}
\bibpunct{(}{)}{;}{a}{}{,}

\newcounter{number}
\newcommand\hii{H$\;${\small\Roman{number}}\ }

\newcommand\arcsec{\mbox{$^{\prime\prime}$}}%
\newcommand\eg{\emph{e.g.\/}}
\newcommand\arcdeg{\mbox{$^\circ$}}%
\newcommand\micron{\mbox{$\mu$m}}%
\newcommand{\dif}{\mathrm{d}}

\TitreGlobal{Sky Polarisation at Far-Infrared to Radio Wavelengths}
\Editors{F. Boulanger, M.-A. Miville-Desch\^{e}nes (eds)}
\begin{document}

\title{Polarized Emission from Interstellar Dust}%
\author{John E. Vaillancourt}
\address{Enrico Fermi Institute, University of Chicago, Chicago, Illinois, USA\\ \email{johnv@oddjob.uchicago.edu}
}
\runningtitle{J. E. Vaillancourt: Polarized Emission from Interstellar Dust}
%
%
\begin{abstract}

Observations of far-infrared (FIR) and submillimeter (SMM) polarized
emission are used to study magnetic fields and dust grains in dense
regions of the interstellar medium (ISM\@).  These observations place
constraints on models of molecular clouds, star-formation, grain
alignment mechanisms, and grain size, shape, and composition.
The FIR/SMM polarization is strongly dependent on
wavelength.  We have attributed this wavelength dependence to sampling
different grain populations at different temperatures.  To date, most
observations of polarized emission have been in
the densest regions of the ISM\@.  Extending these observations to
regions of the diffuse ISM, and to microwave frequencies, will provide
additional tests of grain and alignment models.

An understanding of polarized microwave emission from dust is key to
an accurate measurement of the polarization of the cosmic microwave
background. The microwave polarization spectrum will put limits on the
contributions to polarized emission from spinning dust and vibrating
magnetic dust.

\end{abstract}
\maketitle

\section{Introduction}

The dominant source of Galactic emission at far-infrared (FIR) and
submillimeter (SMM) wavelengths ($\lambda\sim 50\,\micron$ -- 1\,mm,
$\nu\sim 300$ -- 6000 GHz) is thermal emission from interstellar dust
at temperatures of 10 -- 100 K\@.  In dense regions of the
interstellar medium (ISM) such as molecular clouds, this emission has
measurable polarization at almost every point
(\eg\ \citealt{archive,dotson06}).  This polarization is due to the
alignment of the dust grains with interstellar magnetic fields
\citep{dg51}. As a result of the magnetic alignment mechanism,
polarimetry allows detailed studies of magnetic fields
(\eg\ \citealt{chussth,mwf01,crutcher}).

Polarimetry can also be used to place constraints on grain properties,
cloud environments, and grain alignment efficiency
(\citealt{pspec,whittet01}; \citealt*{lgm97}).  Observations at
multiple wavelengths are especially useful for such studies.  At
optical and near-infrared wavelengths, spectropolarimetry puts
constraints on grain size, shape, and composition
\citep{aitken96,whittet04,martin}.  In molecular clouds, the FIR/SMM
polarization spectrum is observed to fall from $60\,\micron$ to
$350\,\micron$ and rise again to $1300\,\micron$.  We attribute this
spectrum to correlations between temperature and alignment efficiency
\citep{pspec,mythesis}.

Further tests of grain and alignment models require that measurements
of the polarization spectrum be extended to more diffuse regions of
the ISM and to longer wavelengths.  Diffuse infrared cirrus clouds
present simpler physical environments than dense star-forming
molecular clouds.  Long wavelength observations ($\lambda \geqslant
850\,\micron$) will determine the behavior of the polarization
spectrum where thermal emission from dust no longer dominates the
total spectral energy distribution.  Experiments such as \emph{Planck}
\citep{tauber04} and the \emph{Wilkinson Microwave Anisotropy Probe}
(\emph{WMAP}; \citealt{hinshaw03}) are designed to make all-sky
polarization measurements at these wavelengths. 

Dust may also be a major contributor to the observed Galactic emission
at microwave frequencies ($\lambda \sim 3$ -- 300 mm).  This dust
emission will be dominated by electric dipole \citep{dl98b} or
magnetic dipole \citep{dl99} radiation rather than thermal radiation.
As a result, Galactic dust emission may be an important foreground
contributor to studies of the cosmic microwave background (CMB\@).
Extension of the polarization spectrum of Galactic clouds from the SMM
to the microwave will be key to accurately removing this polarized
foreground from the CMB\@.

\section{Polarization by Absorption and Emission} \label{sec-poln}

\subsection{Polarization by Absorption}

Polarization of starlight at ultraviolet (UV), visible (Vis), and
near-infrared (IR) wavelengths is due to selective extinction by
aspherical grains which have been aligned by a local magnetic field
\citep{dg51}.  Grain alignment generally occurs in two steps: 1) the
grain axis with the largest moment of inertia becomes aligned with the
spin (angular momentum) axis and 2) the spin axis then becomes aligned
with the local magnetic field
(\eg\ \citealt{lazarian03,whittetbook,roberge04,lazcho_conf}).  While
grain dynamics are certainly more complex than this picture
\citep{lazarian04}, and other alignment mechanisms (\eg\ mechanical
alignment; \citealt{gold52}) may be important in some regions
\citep{rhh88,lazarian03}, this general alignment picture
will suffice for the discussion of polarimetry presented here.

Since starlight is preferentially absorbed along the long axis of the
grain, the net polarization will be parallel to the magnetic field.
Observations of starlight polarization by absorption have proven to be
a valuable tool for tracing the magnetic field structure in diffuse
regions of the interstellar medium (\eg\ \citealt{heilescat};
\citealt*{berdyugin04}; and references therein).  However, at high
extinctions, UV/Vis/IR light is completely absorbed.  Even at moderate
extinctions ($A_V \gtrsim 1$\,magnitude) polarization by absorption is
not a reliable tracer of the magnetic field due to the drop in grain
alignment efficiency (\S\ref{sec-limits};
\citealt{goodman95,arce,bastien}).

\subsection{Polarization by Emission}

\begin{figure}
  \includegraphics[width=2.4in]{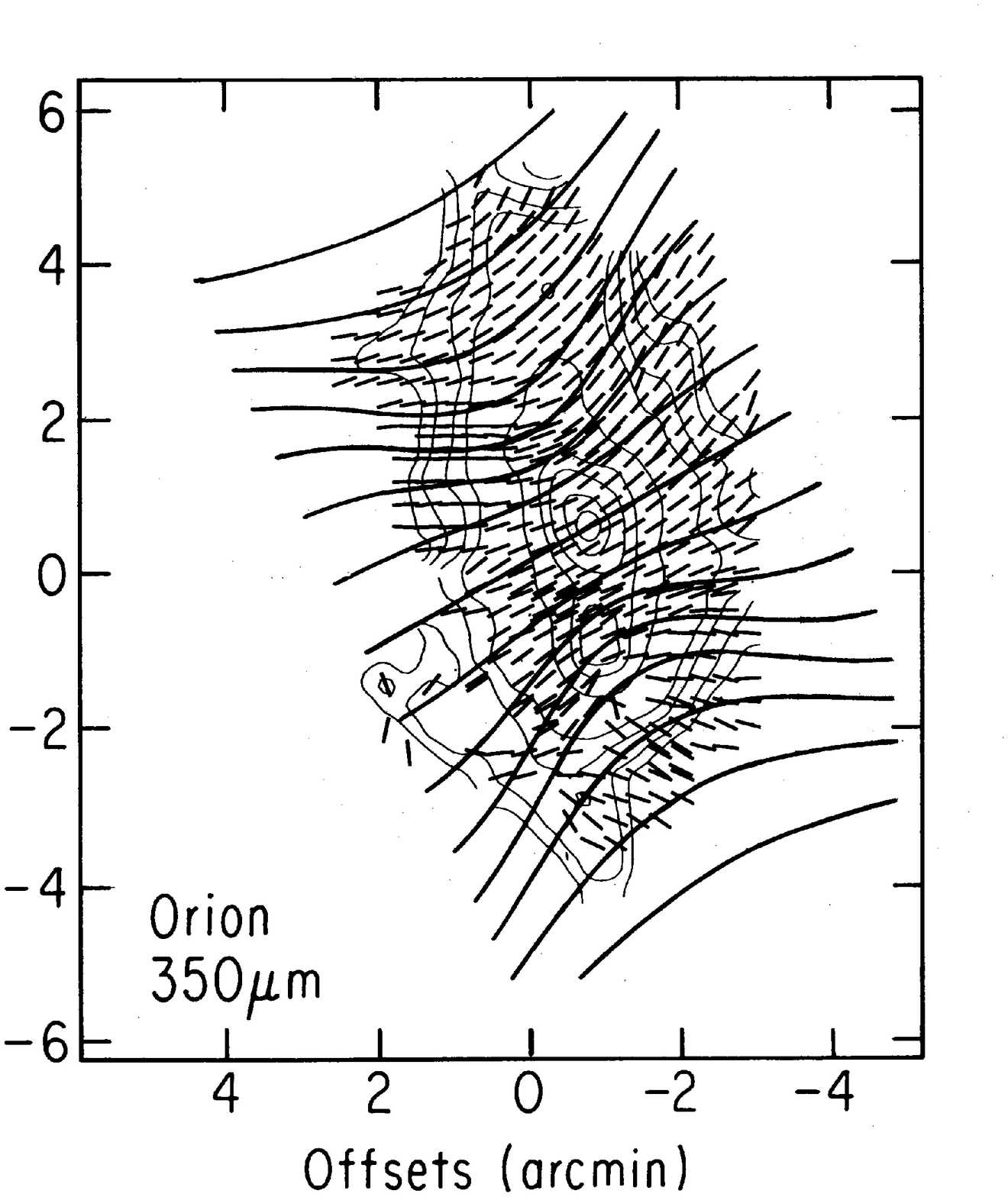}
  \includegraphics[width=2.4in]{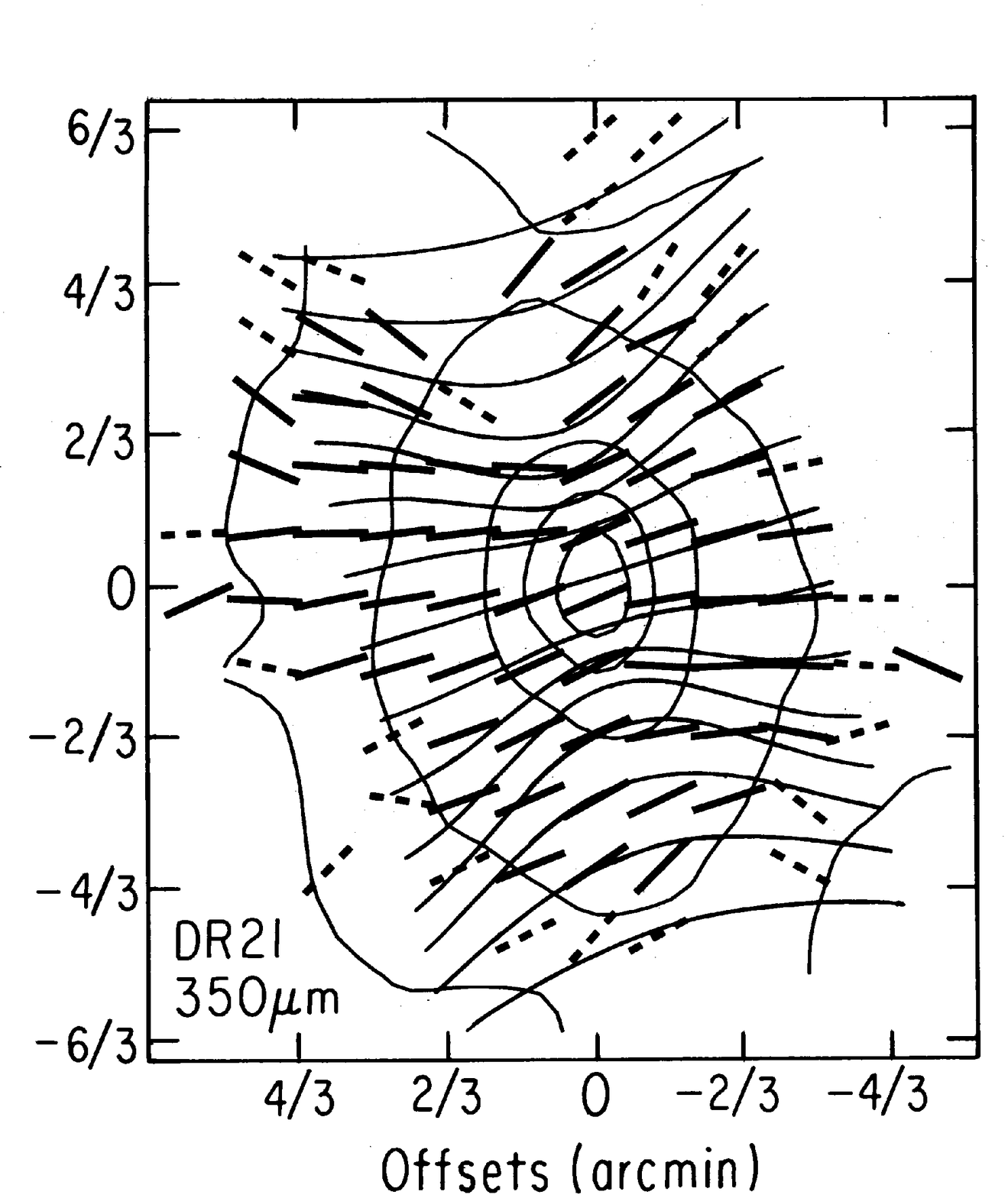}
  \caption{Polarization of the Orion Molecular Cloud (left;
    \citealt{houde04}) and DR21 (right; \citealt{dotson06}) at
    $350\,\micron$. Polarization vectors (small solid and dashed
    lines) are all drawn the same size and are rotated 90 degrees to
    show the inferred magnetic field direction.  The curves show a
    possible model for the magnetic field. For DR21 the curves are
    scaled down by a factor of 3 to account for the relative distances
    of the two clouds. \label{fig-omc1}}
\end{figure}

The net polarization of radiation emitted from dust grains is parallel
to the long axis of the grain, perpendicular to the aligning magnetic
field.  Figure \ref{fig-omc1} shows an example of magnetic fields in
two molecular clouds inferred from emission polarimetry at SMM
wavelengths.

\begin{figure}
\center{\includegraphics[width=4.in]{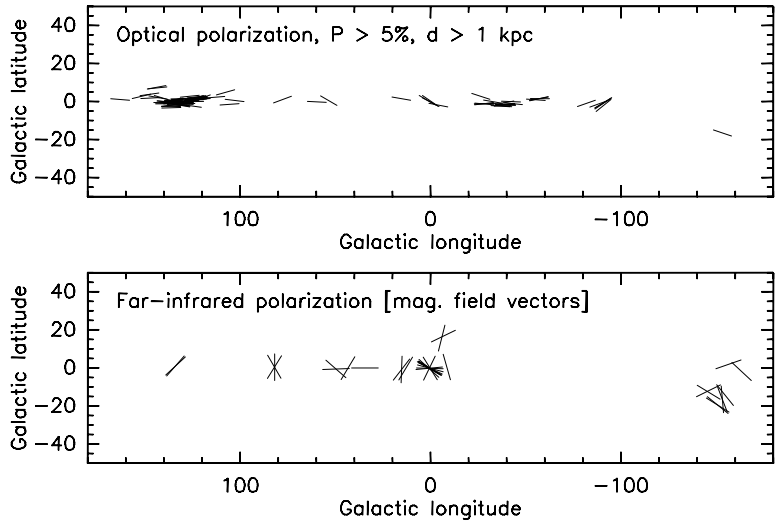}}
	\caption{Top: Galactic polarization measurements of the
          diffuse ISM at optical wavelengths
          \citep{heilescat}. Bottom: Mean field direction for 27
          Galactic molecular clouds observed in the far-infrared and
          submillimeter (\citealt{tenerife}; courtesy
          C. D. Dowell). \label{fig-galpol}}
\end{figure}

Instrument sensitivity, as well as atmospheric absorption and
stability, set a lower limit on the column densities which can be
traced using emission polarimetry.  At present the best submillimeter
instruments require $A_V\gtrsim 10$ -- 20 \citep{sharp}. As a result
polarimetry using emission and absorption trace magnetic fields in
different domains of the ISM\@. Figure \ref{fig-galpol} shows that the
Galactic magnetic field can be quite different in these regions.
Optical polarimetry traces a field which is in general parallel to the
Galactic plane, whereas the FIR polarization vectors appear to be
oriented randomly.  While polarimetry in dense clouds clearly traces
magnetic fields (\S\ref{sec-pflux}) they do not exhibit a clear link
to the large-scale Galactic field.

\section{Grain Alignment} \label{sec-align}

\subsection{Radiative Torques}

The alignment of the rotational and symmetry axes of dust grains with
magnetic fields is opposed by collisions between the grains and gas
molecules.  In order for the grains to become aligned, the timescale
for alignment must be faster than the timescale of the collisional
damping.
\citet{purcell79} showed that this condition is satisfied when the
grains are rotating suprathermally, $E_\mathrm{rot} \gg kT$. He
showed further that torques produced by the formation and subsequent
ejection of H$_2$ molecules from grain surfaces could spin-up grains
to the necessary speeds.

The necessary torques can also be applied by photons
\citep{dolginov72,dolginov76,draine96,draine97}.  Photons will produce
a net torque on irregularly shaped grains because they present
different cross-sections to right- and left-circularly polarized
photons.  Several observations provide support for the existence of
these radiative torques \citep{lazarian03}. Some of these are
discussed further in the remainder of \S\ref{sec-align} and in
\S\ref{sec-spectropol}.

Radiative torques are theoretically more attractive than H$_2$ torques
due to ``thermal flipping'' and ``thermal trapping'' of grains.
\citet{ld97,ld99b} showed that thermal fluctuations in a grain may result
in sudden flips of its angular momentum vector.  Since the location of
H$_2$ formation is dependent on grain surface structure the resulting
torques will change direction when the spin vector flips, causing the
grain to spin-down.  Grains smaller than $1\,\micron$ cannot then 
reach suprathermal velocities \citep{ld99a}.  However, this is not the
case for radiative torques because a grain's helicity is not changed
by a change in the grain's orientation.


\subsection{Limits on Polarization by Absorption} \label{sec-limits}

If all grains are aligned with equal efficiency regardless of their
environment then the number of aligned grains increases at the same
rate as the total number of grains.  One then expects measurements of
polarization by absorption to increase towards regions of higher dust
column density.
\citet{arce} have shown that this expectation holds only for low
extinctions.  At low extinctions interstellar photons can
penetrate the cloud and spin-up the dust grains via radiative torques
but at higher extinctions the cloud is opaque to UV/Vis/IR
photons. Also, as the extinction increases the polarization is further
reduced by scattered photons.  As a result, \citet{arce} conclude that
polarization by absorption cannnot be used to reliably trace magnetic
field structure in regions where $A_V\gtrsim1.3$


If radiative torques are required to spin-up grains to the angular
velocities necessary for alignment then two questions must be answered
in regard to dense molecular clouds where extinctions can be much
larger than $A_V\approx1.3$. What is the mechanism for producing grain
alignment in these clouds? Does the FIR/SMM polarization reliably
trace the orientation of magnetic fields?

\subsection{Radiative Torques in Dense Clouds}

Even in dense clouds, radiative torques can spin grains up to the
suprathermal velocities required for alignment.
While photons from the interstellar radiation field (ISRF) cannot
penetrate deeply into the cloud interior (where $A_V \gg 1.3$),
photons from embedded stars can provide the necessary torques.
\citet{cho05} have shown that radiative torques are still efficient up
to $A_V\sim10$.  Since clouds are likely inhomogeneous, radiation can
leak out to regions of even higher density.  In addition, the
efficiency of radiative torques rises sharply with increasing grain
size \citep{cho05}.  Due to the increased efficiency of grain growth
mechanisms (coagulation, mantle growth) grains tend to be larger in
dense clouds than in the diffuse ISM \citep{whittetbook}.

Stellar photons will also heat the grains, resulting in a correlation
between grain temperature and alignment efficiency which is observable
(\S\ref{sec-spectropol}; \citealt{pspec}).  This correlation will be
strongest along lines of sight to embedded stars and has been observed
towards the \hii regions W3A \citep{w3} and M42 \citep{mythesis}.

\subsection{Polarized Submillimeter Flux} \label{sec-pflux}

\begin{figure}
\begin{center}
  \includegraphics[height=3.5in]{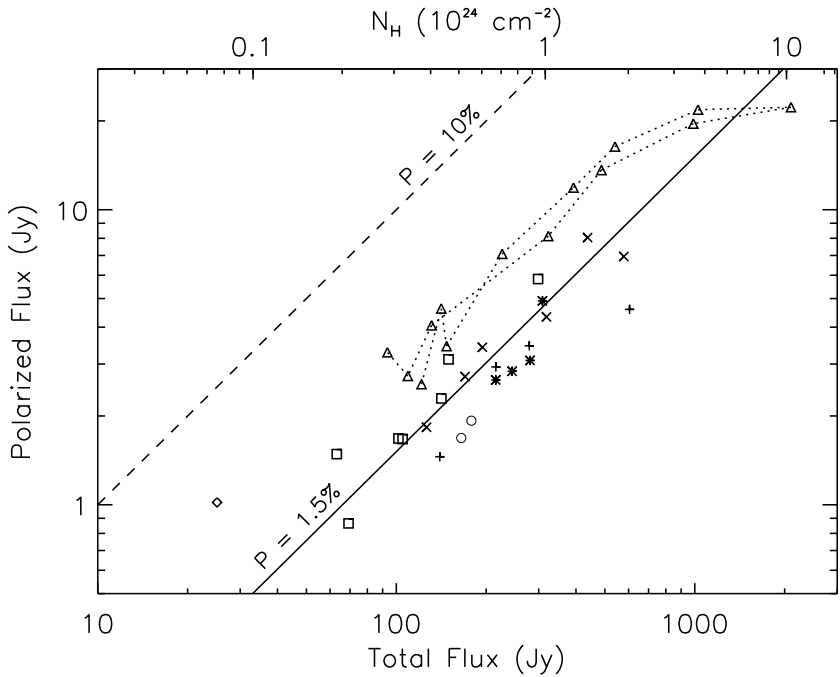}
\end{center}
\caption{Polarized flux vs.\ total flux at $350\,\micron$ for selected
  points in six molecular clouds (Vaillancourt et al. 2006, in
  preparation).  Data points (from \citealt{dotson06}) have
  $P>3\sigma_p$ and were chosen to minimize changes in polarization
  position angle (and therefore changes in the projected magnetic
  field direction) between spatially adjacent points. The solid line is
  the best linear fit to a constant polarization ($P = 1.5\%$) and the
  dashed line is drawn at a constant polarization of 10\%.  A dotted
  line connects data at a constant declination through the core of the
  Orion Molecular Cloud (Figure \ref{fig-omc1}); note the turn over at
  high densities suggesting a loss in grain alignment. Fluxes are
  measured with a $20\arcsec$ beam. The scale at the top is the
  hydrogen column density for a dust temperature of 35\,K;
  $N_\mathrm{H} = 9.2\times 10^{24}$\,cm$^{-2}$ for $\tau(350\micron)
  = 1$ \citep{rhh83}. \label{fig-pflux}}
\end{figure}

The assumption that emission polarimetry traces magnetic fields can be
tested. Again, if all grains are aligned with equal efficiency then
one expects the product of the polarization by emission and column
density to scale with column density. Figure \ref{fig-pflux} plots
this product in a selection of six molecular clouds using the
$350\,\micron$ flux as a tracer of column density. The results are
consistent with the expectations of efficient grain alignment.  This
trend becomes shallower for some of the highest density clouds (Orion;
dotted line in Figure \ref{fig-pflux}) consistent with the idea that
grain alignment is less efficient at higher densities.

These results are more difficult to interpret than for the
polarization by absorption case because the SMM flux is only a good
tracer of dust column density if grains are either isothermal or warm
enough so that the observations are made at wavelengths on the
Rayleigh-Jeans tail. Neither of these conditions is necessarily true
in molecular clouds.

Additionally, the polarization projected onto the plane of the sky
allows only a lower limit to be placed on the grain alignment
efficiency.  While this is also true in the case of absorption, the
projected magnetic field in molecular clouds can change more rapidly
than in the diffuse ISM (\ie\ on angular scales approaching the
FIR/SMM resolution).  An effort has been made to minimize this problem
in Figure \ref{fig-pflux} by avoiding regions where the polarization
position angle changes by more than 10 degrees between spatially
adjacent points.  Even with this position angle cut, the drop in
polarization seen in Orion is partly due to magnetic field structure on
scales below the $20\arcsec$ spatial resolution \citep{rao98}.

\section{Spectropolarimetry} \label{sec-spectropol}

In both the cases of polarization by absorption and polarization by
emission, multi-wavelength polarimetry has been used to study magnetic
fields and dust grain properties.  However, it is in studies of the
dust where spectropolarimetry is truly invaluable (see reviews by
\citealt{martin,whittetbook,whittet04}).  We know that large grains
(radii $\gtrsim0.1\,\micron$) are more efficient polarizers than small
grains (radii $\lesssim0.01\,\micron$) because the polarization
follows the extinction at infrared wavelengths, but not ultraviolet
wavelengths.  We know that silicate grains are better aligned than
graphite grains because the spectral features of the O-Si bond are
polarized, but carbon features are not. From the shape of the spectrum
we know that the aligned grains are oblate (disc-like) rather than
prolate (needle-like) \citep{drainelee84,leedraine85,shape}.


\subsection{Measured FIR/SMM Polarization Spectra}

\begin{figure}
\begin{center}
  \includegraphics[height=2.95in]{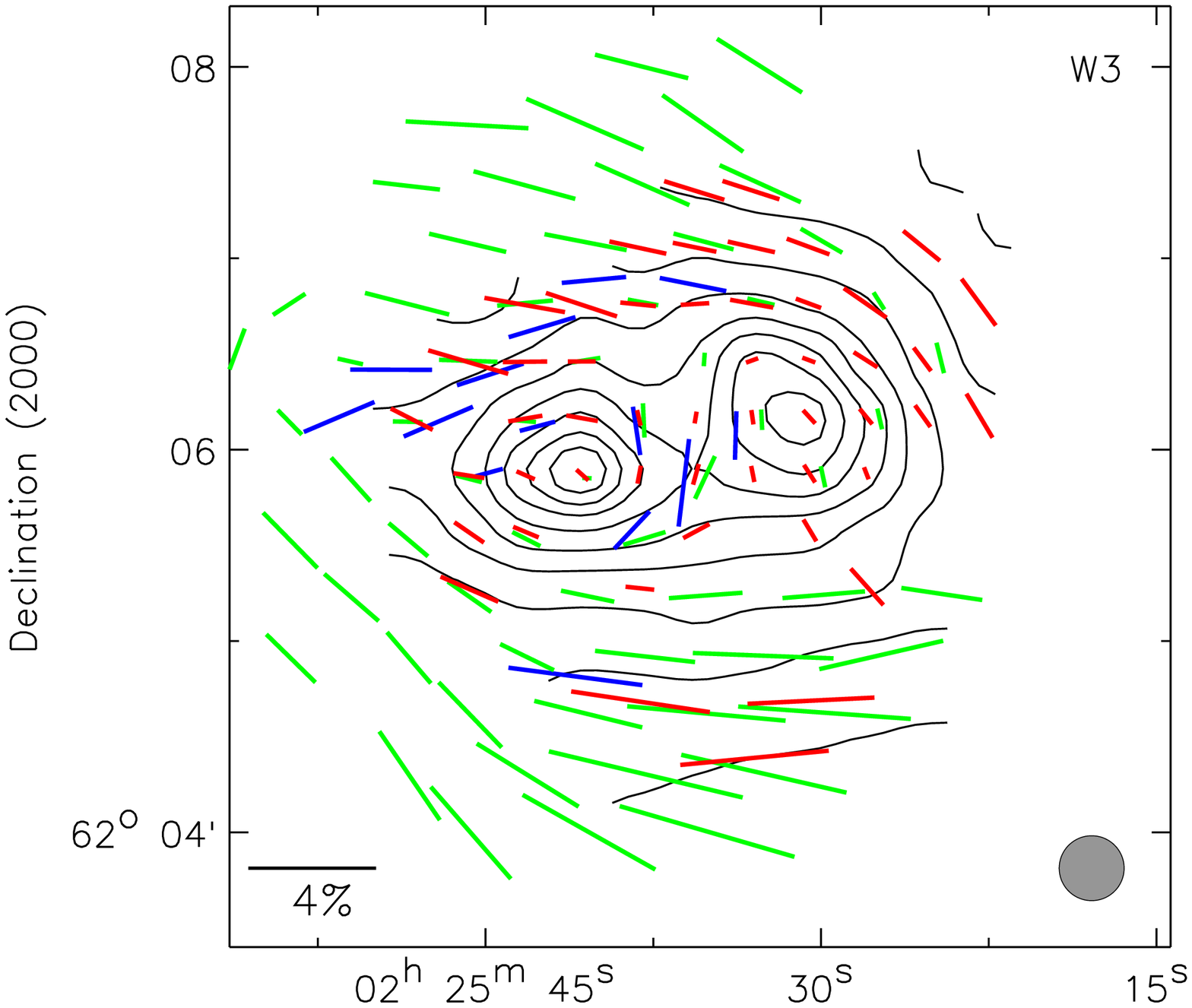}
  \includegraphics[height=2.95in]{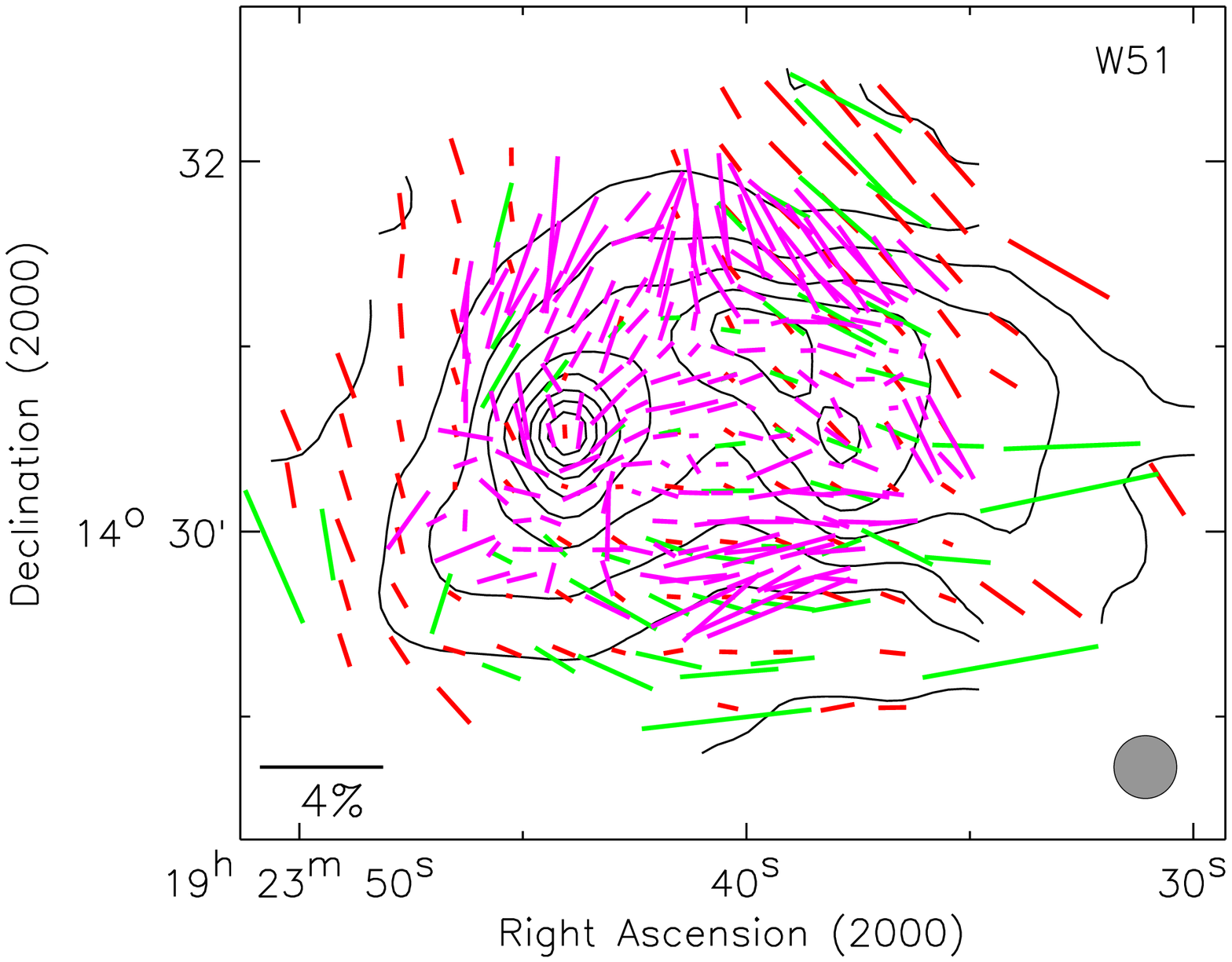}
\end{center}
\caption{Top: Polarization of W3 measured at $60\,\micron$ (blue),
  $100\,\micron$ (green), and $350\,\micron$ (red). Notice the
  clockwise rotation of the vectors from 60 to 100 to 350 $\micron$,
  signifying a change in the projected magnetic field along the line
  of sight, presumably due to differential rotation of the inner,
  warmer dust with respect to the outer, cooler dust \citep{w3}.
  Bottom: Polarization of W51 at $100\,\micron$ (green,
  \citealt{archive}), $350\,\micron$ (red, \citealt{dotson06}), and
  $850\,\micron$ (magenta, \citealt{w51scuba}). Shaded circles show
  the $20\arcsec$ beam used to make the $350\,\micron$ measurements
  and flux contours.\label{fig-maps1}}
\end{figure}

Figure \ref{fig-maps1} shows two examples of multi-wavelength
polarimetry of molecular clouds in the FIR/SMM\@.
In comparing the percent polarization at different wavelengths one
must consider that the changes can be due to changes in the strength
or projection of the magnetic field, the efficiency of the alignment
process, and the emissivity of the grains.  All of these parameters
may vary with position within the cloud.  One may limit the effect of
the magnetic field on these changes by avoiding regions where the
position angle changes greatly from nearby points, or changes
with wavelength.  Such position angle changes can only be due to
changes in the magnetic field.


\begin{figure}
  \includegraphics[width=2.4in]{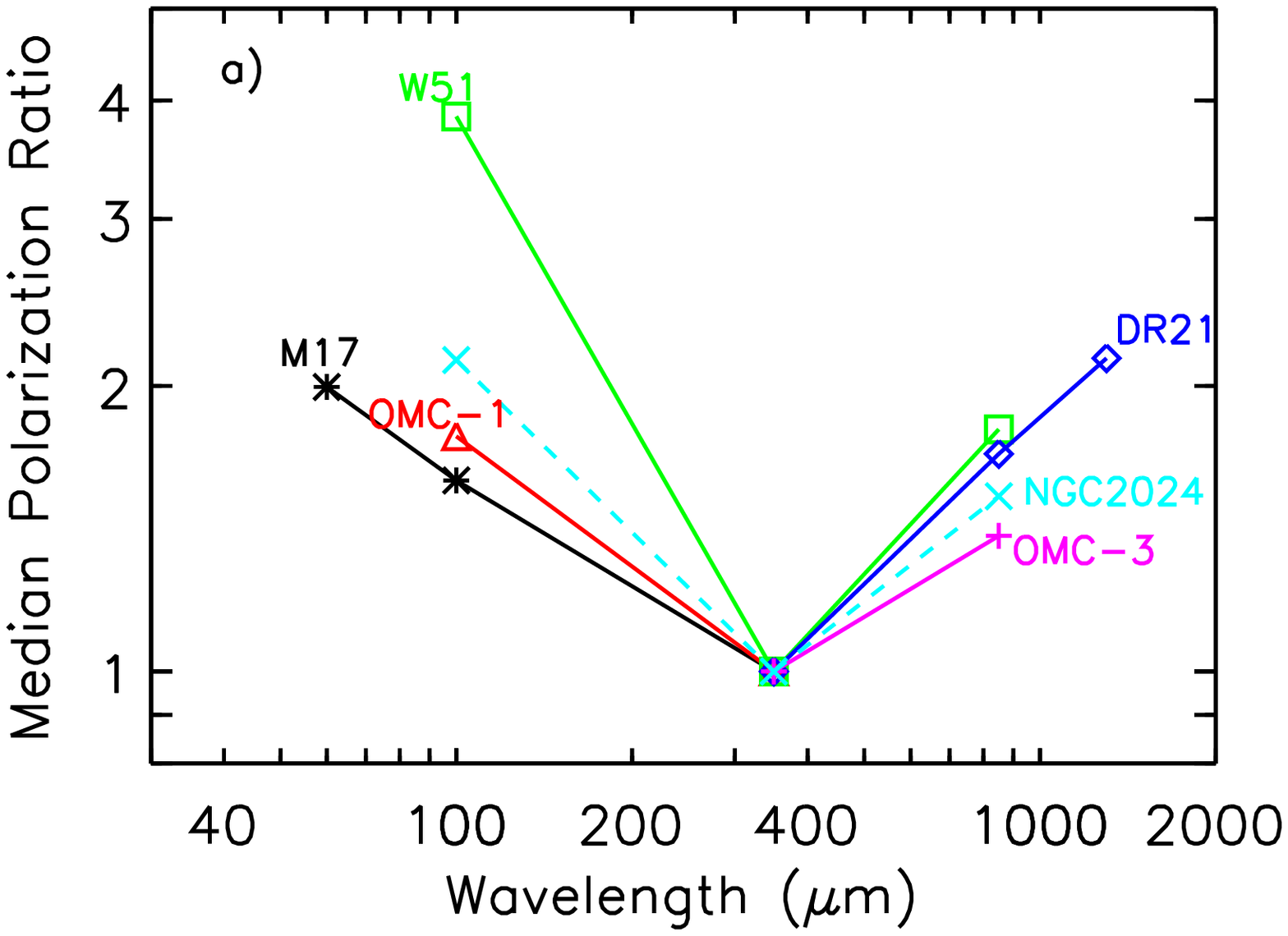}
  \quad
  \includegraphics[width=2.4in]{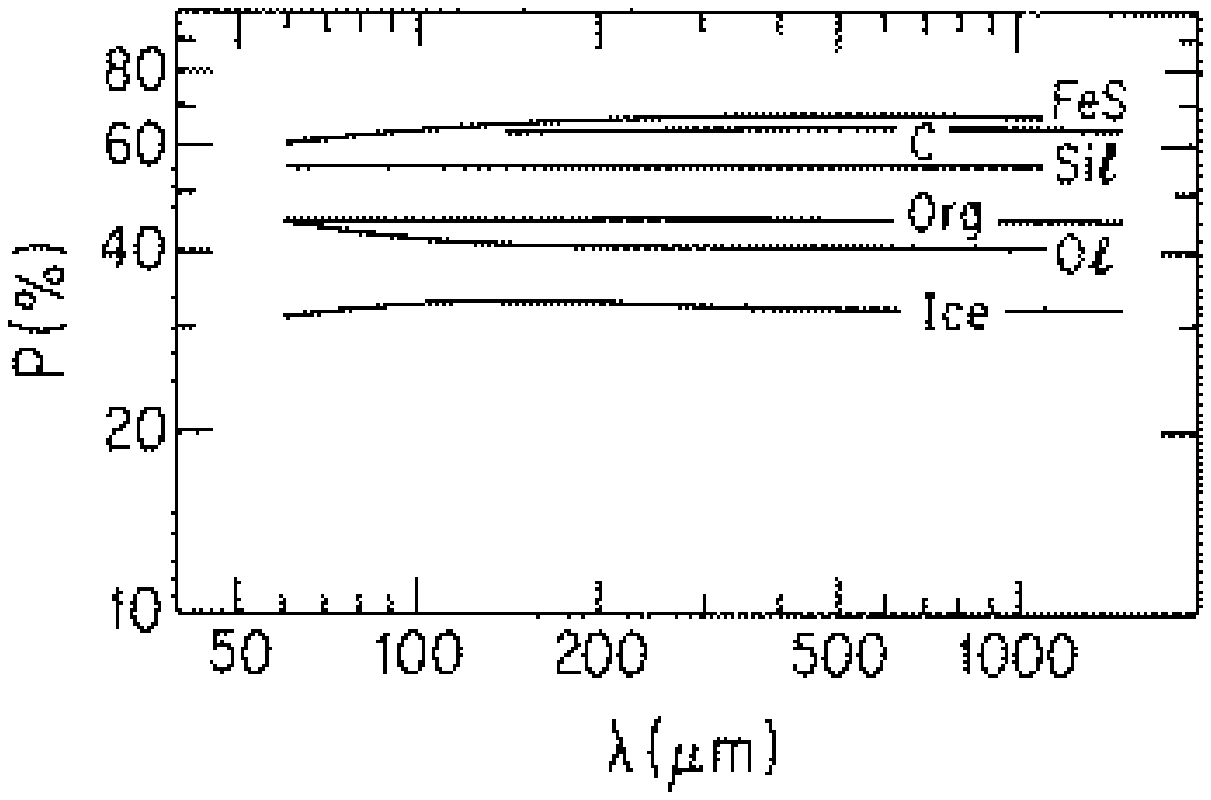}
  \caption{Measured and predicted far-infrared/submillimeter
    polarization spectra for interstellar dust. a) Measured
    polarization spectra for molecular cloud envelopes, normalized at
    $350\,\micron$ (\citealt{mythesis,brenda-pspec}; W51
    $850\,\micron$ data from \citealt{w51scuba}). b) Estimated
    polarization spectra from individual grain species assuming
    perfect alignment, a single temperature, and a constant grain size
    and shape \citep{pspec}.\label{fig-pspec}}
\end{figure}

Figure \ref{fig-pspec}a shows polarization spectra in 6 molecular
cloud envelopes. (Envelopes are defined as regions away from dense
cores where opacity may affect the spectrum.)  The spectrum falls from
$60\,\micron$ to $350\,\micron$ and then rises from $350\,\micron$ to
$1300\,\micron$. The uncertainties in these comparisons are large,
and data exist both short- and long-ward of $350\,\micron$ for only
two of the clouds shown here (W51 and NGC2024).  However, the trend is
observed in multiple clouds, and in clouds where data are available
both short- and long-ward of the $350\,\micron$ minimum.
Continued observations of these and similar clouds at FIR/SMM
wavelengths are needed to further confirm these results
(\citealt{hale,sharp,sharp2}; \citealt*{scuba2}).

\subsection {Models of the Polarization Spectrum}

The wavelength dependence of polarization for specific types of grains
can be calculated assuming uniform magnetic fields.
Using the known dielectric properties of grain materials,
\citet{pspec} found that an ideal cloud composed of identical grains
at a single temperature would produce a polarization nearly
independent of wavelength for several different grain materials
(Figure \ref{fig-pspec}b).

Molecular clouds are certainly not ideal.  This is confirmed by the
polarization spectrum in Figure \ref{fig-pspec}a which clearly does
not reproduce the prediction in Figure \ref{fig-pspec}b. Real clouds
are likely composed of multiple grain species with a distribution of
temperatures.  The total polarization is found by summing the
polarized flux from each component \citep{pspec}:
\begin{equation}
P_\mathrm{tot}(\lambda) = \frac{1}{F_\mathrm{tot}(\lambda)} \sum_i p_i F_i(\lambda).
\label{eq:pmix1}
\end{equation}
where $F_i(\lambda)$ is the flux from species $i$,
$F_\mathrm{tot}(\lambda) = \sum_i F_i(\lambda)$, and $p_i$ is the
wavelength independent polarization of species $i$ from Figure
\ref{fig-pspec}b.

Consider a cloud composed of dust components at two temperatures, each
with a different polarization efficiency.  A wavelength dependent
polarization spectrum then requires that two conditions be true:
\begin{equation}
p_1 \lessgtr p_2 \qquad \mathrm{and} \qquad
\frac{1}{F_1} \frac{\dif F_1}{\dif\lambda} >
\frac{1}{F_2} \frac{\dif F_2}{\dif\lambda} 
\label{eq-preq}
\end{equation}
where the upper inequality holds for a falling spectrum ($\dif
P_\mathrm{tot}/\dif\lambda < 0$) and the lower inequality for a rising
spectrum ($\dif P_\mathrm{tot}/\dif\lambda > 0$).  If the total emitted
dust flux is modeled as an emissivity-modified blackbody with spectral
index $\beta$ [$F_i(\nu) \propto \nu^{\beta_i} B_\nu(T_i)$] then the
second condition is equivalent to
\begin{eqnarray}
T_1 < T_2 & \quad \mathrm{if} \quad & \beta_1 = \beta_2 \quad \mathrm{or}\\
\beta_1 < \beta_2 & \quad \mathrm{if} \quad & T_1 = T_2.
\label{eq-pmix2}
\end{eqnarray}

The falling spectra between 60 -- $350\,\micron$ in Figure
\ref{fig-pspec}a are consistent with a model in which the warmer grains
are better aligned than the cold grains \citep{pspec,mythesis}.  Such
regions may occur naturally in molecular clouds.  Grains near embedded
stars will be both warmed by stellar photons and be spun-up to
suprathermal velocities by radiative torques, thereby increasing the
grain alignment efficiency.
Grains further from the star, or in optically thick clumps where
stellar photons cannot penetrate, will be cooler and less affected by
radiative torques.  \citet{mythesis} has shown that the spectral
energy distribution and polarization spectra in the Orion nebula (M42)
\hii region are consistent with this model.

The rising spectra longward of $350\,\micron$ require another dust
component with either a colder temperature or a lower spectral index
than the dust in the dense clumps.  Such dust may occur on the surface
layers of molecular clouds where photons from the ISRF can impart
radiative torques.

\section{Diffuse Clouds} \label{sec-diffuse}

Diffuse low density clouds present a simpler case for studies of
interstellar dust grains than molecular clouds.  In clouds transparent
to starlight all grains are exposed to the same, nearly isotropic,
radiation source.  Such clouds would represent the ideal case of
Figure \ref{fig-pspec}b except that grain species with different
emissivities will have different equilibrium temperatures.  The
diffuse infrared cirrus clouds observed at high Galactic latitude by
the \emph{Infrared Astronomical Satellite} (\emph{IRAS};
\citealt{low84}) are the best observational examples of ideal
clouds. These clouds are distributed over most of the sky and are
bright enough (10--100 MJy/sr at $\lambda=100$ -- 200 \micron) to be
detectable with the next generation of airborne photometers
\citep{hawc} and polarimeters \citep{hale}.

The FIR/SMM spectral energy distribution of the Galaxy can be modeled
by two dust components with mean temperatures of 9.4\,K and 16\,K and
spectral indices of 1.7 and 2.7, respectively \citep*{fds99}.
Measurements of the optical properties of graphites and silicates
\citep{drainelee84,pollack94} led \citet{fds99} to tentatively
identify silicate grains with the $\beta=1.7$ component and
carbonaceous grains with the $\beta=2.7$ component.  Silicate grains
reach colder equilibrium temperatures than graphites when both are
exposed to the ISRF.  (\citeauthor{drainelee84} estimated higher
temperatures than those of \citeauthor{fds99})
With silicate colder than graphite, and silicate grains more polarized
than carbon grains \citep{whittet04} equations
(\ref{eq-preq})--(\ref{eq-pmix2}) predict that polarized emission from
diffuse cirrus clouds will increase with wavelength
(\S\ref{sec-uwavemodel}; \citealt{aod}).  Tests of this prediction
require observations at FIR/SMM wavelengths. Polarimetric observations
from the \emph{Stratospheric Observatory for Infrared Astronomy
  (SOFIA)} will cover the 60 -- 215 $\micron$ range \citep{hale}.
Submillimeter wavelengths can be covered by
\emph{Planck} at 350 -- 1400 $\micron$ \citep{tauber04},
\emph{Archeops} at $850\,\micron$ \citep{ponthieu,benoit04}, and
\emph{BLAST} at 250 -- 500 $\micron$ \citep{netterfield}.

\section{Microwave Emission} \label{sec-uwave}


\subsection{Total Microwave Flux Spectra}

Several studies covering large areas on the sky in the 10 -- 90 GHz
range have reported microwave emission which is highly correlated with
thermal dust emission \citep{kogut96,doc97,leitch97}.  The microwave
flux is larger than that expected from a combination of only thermal
dust, free-free emission, and synchrotron emission
\citep{doc99,dl98a},
suggesting the existence of a new emission component in this spectral
region.
The strong correlations between this excess microwave flux and dust
emission at shorter far-infrared wavelengths have led to two models in
which dust is the source of the excess flux (see review by
\citealt{lazfink}).

In the first model, small grains (radii $\lesssim0.001\,\micron$) have
dipole moments, spin rapidly, and emit electric dipole radiation at
microwave frequencies \citep{dl98a,dl98b}.  These so-called
``spinning-dust'' grains are the same small grains responsible for
extinction at ultraviolet wavelengths in diffuse regions of the ISM\@.
In the second mechanism, thermal fluctuations in large grains (radii
$\gtrsim0.1\,\micron$), the same grains responsible for emission in
the far-infrared, lead to fluctuations of the grains' magnetization.
The result is magnetic dipole emission from ``vibrating-dust'' grains
at microwave frequencies \citep{dl99}.  

\begin{figure}
	\includegraphics[width=4.8in]{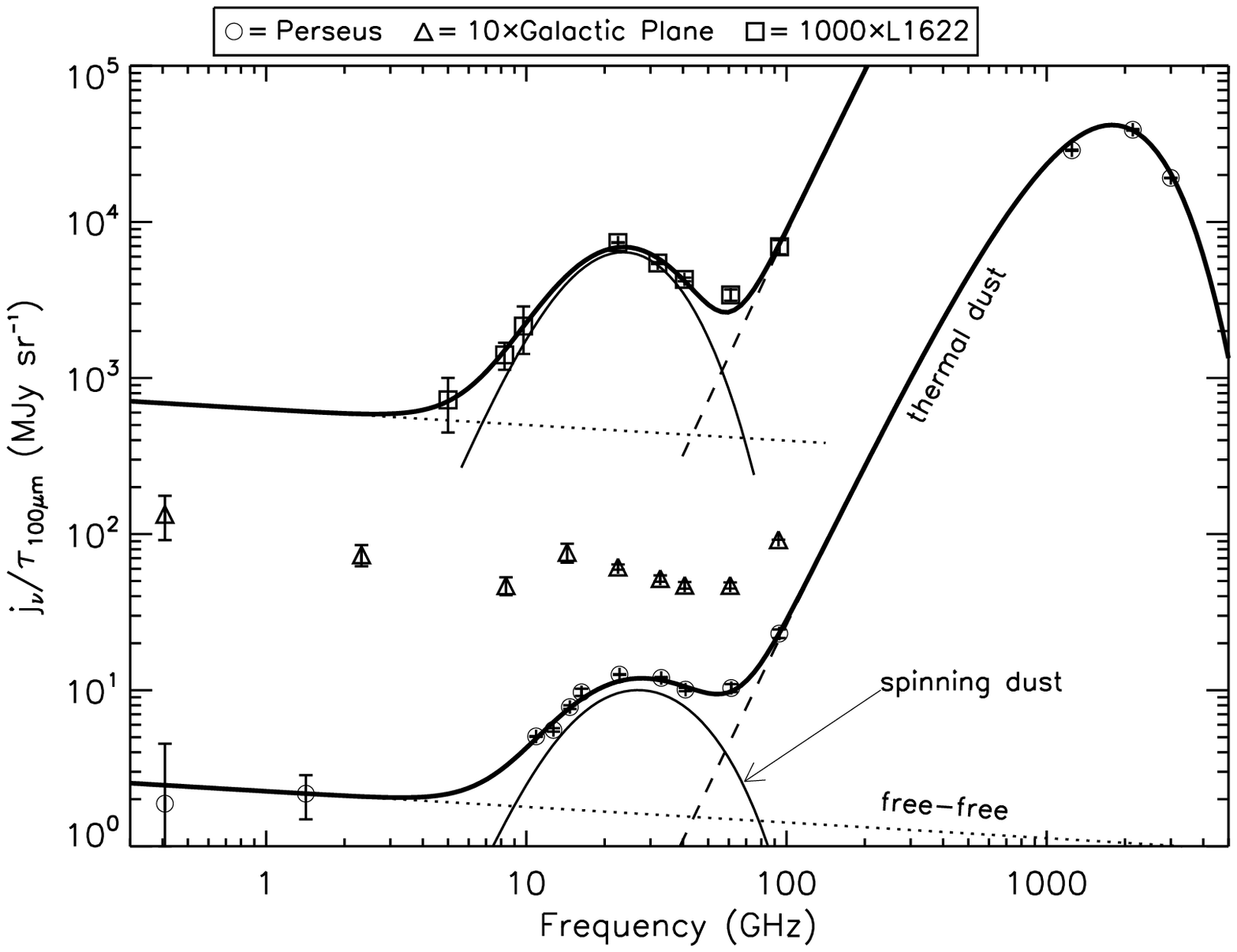}
	\caption{Three examples of spectral energy distributions with
          excess microwave emission.  The flux is in units of emission
          per $100\,\micron$ optical depth
          (\eg\ \citealt{fink04}). Circles: G159.6-18.5 in the Perseus
          molecular cloud \citep{watson05}.  Triangles: a region of
          diffuse emission in the Galactic plane ($|b| < 4\arcdeg$,
          $l\sim 45\arcdeg$), $\times 10$ \citep*{flm04}.  Squares: the
          dark cloud Lynds 1622, $\times 1000$
          \citep{fink04,aod}. Also shown are possible total emission
          models for the Perseus and Lynds 1622 clouds (thick solid
          lines).  Both models include thermal dust emission (dashed),
          free-free emission (dotted), and spinning dust emission
          (thin solid lines).  The Galactic plane emission can also be
           modeled with a combination of these mechanisms and
          synchrotron emission \citep{flm04}. \label{fig-datased}}
\end{figure}

Figure \ref{fig-datased} shows three examples of observations in which
the microwave emission cannot be modeled by only free-free,
synchrotron, and thermal dust emission.  The excess fluxes are well
fitted to electric dipole emission from spinning dust.  However, the
model uncertainties are large enough that the flux data are not
sufficient to rule out vibrating magnetic dust as a significant
component \citep*{aod,flm04}.


Regardless of the actual carrier of the microwave excess, it is a
large contributor to the total flux at 5 -- 50 GHz and may have a
significant impact on measurements of the cosmic microwave background.
Models consisting of only synchrotron or free-free emission below
100\,GHz can be ruled out for the regions shown in Figure
\ref{fig-datased}.
The rising spectra in the 5 -- 20 GHz range are inconsistent with the
expected spectral shape of these two mechanisms.

\subsection{Microwave Polarization Spectra}

Extension of the polarization spectrum to microwave wavelengths is a
matter of adding the polarized contributions from free-free,
synchrotron, and spinning and/or vibrating dust to equation
(\ref{eq:pmix1}).  One can expect that the small grains responsible
for electric dipole emission will be poorly aligned with magnetic
fields \citep{martin,whittet04}. Consequently, they will emit very
little polarized radiation.  \cite{ld00} provide an upper limit on the
alignment of spinning dust grains.  They predict spinning dust will be
polarized no more than 7\% at 2\,GHz, dropping to $< 0.5\%$ at
frequencies above 30\,GHz.  The polarization position angle for
spinning grains will be perpendicular to the aligning field, parallel
to the polarized emission from thermal dust.

However, magnetic dipole emission from vibrating grains should be
highly polarized.  \citet{dl99} predict a polarization level as high
as 40\% for grains with single magnetic domains.  Additionally, for
certain grain shapes, the polarization position angle may flip by 90
degrees as the frequency changes.  At high frequencies the
polarization will be perpendicular to the magnetic field, as it is for
the cases of spinning and thermal dust.  At low frequencies the
polarization will be parallel to the field.

The different polarization spectra for spinning and vibrating dust
should allow for observational tests of these mechanisms where the
flux data alone did not.  The flip in the polarization position angle
with frequency is the definitive signature of magnetic dipole
emission.  With the exception of frequencies near the position angle
flip, magnetic dipole radiation from all single-domain grains is
highly polarized, whereas emission from spinning grains is mostly
unpolarized above 30\,GHz ($\lambda<10$\,mm).

\subsection{Flux and Polarization Spectra of Diffuse Clouds} \label{sec-uwavemodel}

\begin{figure}
	\includegraphics[width=4.8in]{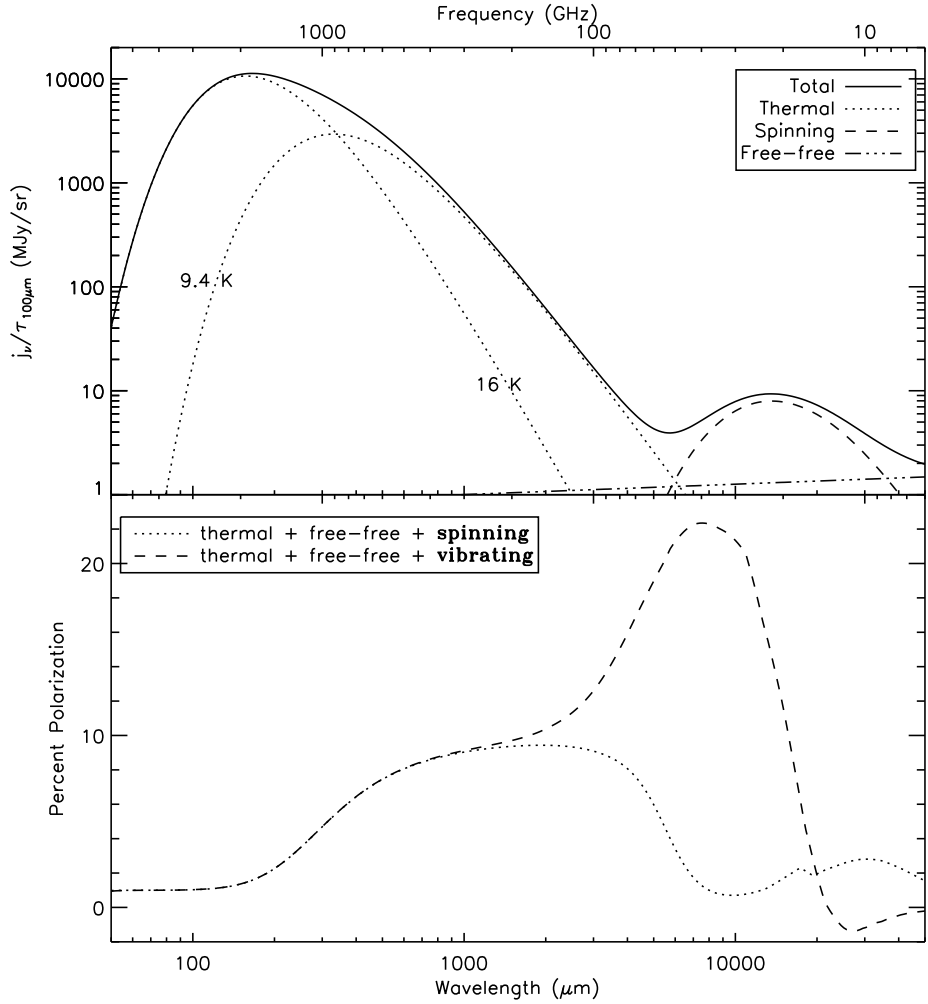}
	\caption{Top: Far-infrared -- microwave spectral energy
          distribution of a hypothetical diffuse cloud.  Flux
          components include thermal emission from dust at 9.4\,K and
          16\,K \citep{fds99}, electric dipole emission from spinning
          dust (cold neutral medium prediction of \citealt{dl98b}),
          and free-free emission. Bottom: Predicted polarization
          spectra for emission including upper limits for spinning
          dust and vibrating magnetic dust.  Thermal dust is assumed
          polarized at the 10\% level for the 9.4\,K component and
          unpolarized for the 16\,K component.  Spinning dust
          polarization is taken from the resonance relaxation model of
          \citet{ld00}.  The vibrating dust SED and polarization
          curves are for the Fe-rich material X4 of \citet{dl99}.  In
          order to calculate the polarized flux from magnetic dust the
          SED peak of magnetic dust (not shown in top panel) is scaled
          to match the peak of the spinning dust SED.  The
          polarization is for single-domain grains with axes ratios
          1:1.5:2 \citep{dl99}.  Free-free emission is assumed to be
          unpolarized. \label{fig-model}}
\end{figure}

Consider a diffuse cloud composed of thermal dust at two temperatures
(9 and 16 K), spinning dust\footnote{Spinning dust predictions were
  downloaded from
  http://www.astro.princeton.edu/~draine/dust/dust.html}, and free
electrons.  The emission spectrum of such a cloud will be dominated by
thermal dust at $\lambda \sim 50\,\micron$ to a few millimeters
($\nu\sim$100 -- 6000 GHz).  At longer wavelengths it may be dominated
by electric dipole emission from spinning dust, and at yet longer
wavelengths by free-free and/or synchrotron emission.  Figure
\ref{fig-model} shows a schematic spectrum of a hypothetical cloud
where synchrotron is subdominant in the entire $50\,\micron$ -- 5\,cm
range.

Figure \ref{fig-model} also shows a predicted polarization spectrum
for such a cloud. If we assume that the warm thermal dust component
is unpolarized while the cold component is polarized ($P\sim10\%$ in
Figure \ref{fig-model}) then the polarization spectrum will rise in
the FIR/SMM (\S\ref{sec-diffuse}).
The polarization drops at $\lambda>1$\,mm ($\nu<300$\,GHz) where the
flux is dominated by unpolarized radiation from spinning dust and
free-free emission.  The polarization may rise for $\lambda >1$\,cm
($\nu<30$\,GHz) if the prediction for the polarization of spinning
dust is correct \citep{ld00}.

If the dominant source of excess microwave emission is magnetic dipole
emission from vibrating dust then the polarization spectrum in Figure
\ref{fig-model} may rise at $\lambda>1$\,mm where this emission
dominates that from thermal dust. In this toy model, the polarization
from these two different mechanisms differs by $\Delta P\sim 20\%$ at
40\,GHz, well within the uncertainties of the current generation of
polarimeters ($\Delta P\lesssim1\%$).  Future polarization data from
\emph{WMAP} \citep{hinshaw03} and \emph{Planck} \citep{tauber04}
should clearly distinguish between spinning and vibrating dust.

\section{Summary}

Dust grains aligned with magnetic fields in the interstellar medium
are responsible for both polarization by absorption at optical
wavelengths and polarization by emission at far-infrared and
submillimeter wavelengths.  Observations of the absorption and
emission trace different regions of the ISM but both are consistent
with the existence of radiative torques which spin grains up to the
suprathermal angular velocities needed for magnetic alignment.

In molecular clouds the FIR/SMM polarization is strongly dependent on
wavelength.  We have attributed this wavelength dependence to sampling
different grain populations at different temperatures.  This
wavelength dependence provides an opportunity to study, in different
physical environments, the grain properties, the alignment mechanism
and its efficiency, and magnetic fields. To date, most observations of
polarized emission have been in
the densest regions of the ISM\@.  Extending these observations to
regions of the diffuse ISM will provide additional tests of grain and
alignment models.  Instruments on board the \emph{Stratospheric
  Observatory for Infrared Astronomy (SOFIA)} will have the
sensitivity to detect the total (HAWC: \citealt{hawc}) and polarized
(Hale: \citealt{hale}) flux in diffuse infrared cirrus clouds.
Instruments, including \emph{Planck} and \emph{WMAP}, designed to
separate polarized Galactic foregrounds from the CMB will be able to
extend the diffuse cloud polarization spectrum to longer SMM and
microwave wavelengths.

The polarization state of the excess microwave emission observed at
frequencies of 10 -- 90 GHz is key to an accurate measurement of
polarization from the cosmic microwave background.  Total power
observations are currently unable to conclusively rule out either of
the two leading candidates for this emission: electric dipole emission
from small spinning grains and magnetic dipole emission from larger
vibrating grains.  Polarization observations made with \emph{WMAP} and
\emph{Planck} will place limits on the emission contributions from
these two sources.

\begin{acknowledgements}
I would like to thank Roger Hildebrand, Alex Lazarian, and Darren
Dowell for valuable comments on drafts of this manuscript.  This work
has been supported in part by NSF grants AST-0204886 and AST-0505124.
\end{acknowledgements}

\bibliography{vaillancourt}

\begin{thebibliography}{70}
\expandafter\ifx\csname natexlab\endcsname\relax\def\natexlab#1{#1}\fi

\bibitem[{{Aitken}(1996)}]{aitken96}
{Aitken}, D.~K. 1996, in ASP Conf.\ Ser.\ 97, Polarimetry of the Interstellar
  Medium, ed. W.~G. Roberge \& D.~C.~B. Whittet, 225

\bibitem[{{Arce} {\etal}(1998){Arce}, {Goodman}, {Bastien}, {Manset}, \&
  {Sumner}}]{arce}
{Arce}, H.~G., {Goodman}, A.~A., {Bastien}, P., {Manset}, N., \& {Sumner}, M.
  1998, \apjl, 499, L93

\bibitem[{Bastien(2006)}]{bastien}
Bastien, P. 2006, this volume

\bibitem[{Bastien {\etal}(2005)Bastien, Jenness, \& Molnar}]{scuba2}
Bastien, P., Jenness, T., \& Molnar, J. 2005, in ASP Conf.\ Ser.\ 343,
  Astronomical Polarimetry --- Current Status and Future Directions, ed.
  A.~Adamson, in press

\bibitem[{{Beno{\^i}t} {\etal}(2004){Beno{\^i}t}, {Ade}, {Amblard}, {Ansari},
  {Aubourg}, {Bargot}, {Bartlett}, {Bernard}, {Bhatia}, {Blanchard}, {Bock},
  {Boscaleri}, {Bouchet}, {Bourrachot}, {Camus}, {Couchot}, {de Bernardis},
  {Delabrouille}, {D{\'e}sert}, {Dor{\'e}}, {Douspis}, {Dumoulin}, {Dupac},
  {Filliatre}, {Fosalba}, {Ganga}, {Gannaway}, {Gautier}, {Giard},
  {Giraud-H{\'e}raud}, {Gispert}, {Guglielmi}, {Hamilton}, {Hanany},
  {Henrot-Versill{\'e}}, {Kaplan}, {Lagache}, {Lamarre}, {Lange},
  {Mac{\'{\i}}as-P{\'e}rez}, {Madet}, {Maffei}, {Magneville}, {Marrone},
  {Masi}, {Mayet}, {Murphy}, {Naraghi}, {Nati}, {Patanchon}, {Perrin}, {Piat},
  {Ponthieu}, {Prunet}, {Puget}, {Renault}, {Rosset}, {Santos}, {Starobinsky},
  {Strukov}, {Sudiwala}, {Teyssier}, {Tristram}, {Tucker}, {Vanel}, {Vibert},
  {Wakui}, \& {Yvon}}]{benoit04}
{Beno{\^i}t}, A., \etal\ 2004, \aap, 424, 571

\bibitem[{{Berdyugin} {\etal}(2004){Berdyugin}, {Piirola}, \&
  {Teerikorpi}}]{berdyugin04}
{Berdyugin}, A., {Piirola}, V., \& {Teerikorpi}, P. 2004, \aap, 424, 873

\bibitem[{{Cho} \& {Lazarian}(2005)}]{cho05}
{Cho}, J. \& {Lazarian}, A. 2005, \apj, 631, 361

\bibitem[{Chrysostomou {\etal}(2002)Chrysostomou, Aitken, Jenness, Davis,
  Hough, Curran, \& Tamura}]{w51scuba}
Chrysostomou, A., Aitken, D.~K., Jenness, T., Davis, C.~J., Hough, J.~H.,
  Curran, R., \& Tamura, M. 2002, \aap, 385, 1014

\bibitem[{Chuss {\etal}(2003)Chuss, Davidson, Dotson, Dowell, Hildebrand,
  Novak, \& Vaillancourt}]{chussth}
Chuss, D.~T., Davidson, J.~A., Dotson, J.~L., Dowell, C.~D., Hildebrand, R.~H.,
  Novak, G., \& Vaillancourt, J.~E. 2003, \apj, 599, 1116

\bibitem[{Crutcher(2006)}]{crutcher}
Crutcher, R.~M. 2006, this volume

\bibitem[{{Davis} \& {Greenstein}(1951)}]{dg51}
{Davis}, L.~J. \& {Greenstein}, J.~L. 1951, \apj, 114, 206

\bibitem[{{de Oliveira-Costa} {\etal}(1997){de Oliveira-Costa}, {Kogut},
  {Devlin}, {Netterfield}, {Page}, \& {Wollack}}]{doc97}
{de Oliveira-Costa}, A., {Kogut}, A., {Devlin}, M.~J., {Netterfield}, C.~B.,
  {Page}, L.~A., \& {Wollack}, E.~J. 1997, \apjl, 482, L17

\bibitem[{{de Oliveira-Costa} {\etal}(1999){de Oliveira-Costa}, {Tegmark},
  {Gutierrez}, {Jones}, {Davies}, {Lasenby}, {Rebolo}, \& {Watson}}]{doc99}
{de Oliveira-Costa}, A., {Tegmark}, M., {Gutierrez}, C.~M., {Jones}, A.~W.,
  {Davies}, R.~D., {Lasenby}, A.~N., {Rebolo}, R., \& {Watson}, R.~A. 1999,
  \apjl, 527, L9

\bibitem[{{Dolginov}(1972)}]{dolginov72}
{Dolginov}, A.~Z. 1972, \apss, 18, 337

\bibitem[{{Dolginov} \& {Mytrophanov}(1976)}]{dolginov76}
{Dolginov}, A.~Z. \& {Mytrophanov}, I.~G. 1976, \apss, 43, 257

\bibitem[{Dotson {\etal}(2000)Dotson, Davidson, Dowell, Schleuning, \&
  Hildebrand}]{archive}
Dotson, J.~L., Davidson, J., Dowell, C.~D., Schleuning, D.~A., \& Hildebrand,
  R.~H. 2000, \apjs, 128, 335

\bibitem[{Dotson {\etal}(2006)Dotson, Dowell, Hildebrand, Kirby, \&
  Vaillancourt}]{dotson06}
Dotson, J.~L., Dowell, C.~D., Hildebrand, R.~H., Kirby, L., \& Vaillancourt,
  J.~E. 2006, in preparation

\bibitem[{Dowell {\etal}(2003)Dowell, Davidson, Dotson, Hildebrand, Novak,
  Rennick, \& Vaillancourt}]{hale}
Dowell, C.~D., Davidson, J.~A., Dotson, J.~L., Hildebrand, R.~H., Novak, G.,
  Rennick, T.~S., \& Vaillancourt, J.~E. 2003, in Proc.\ SPIE 4843, Polarimetry
  in Astronomy, ed. S.~Fineschi, 250

\bibitem[{{Draine} \& {Lazarian}(1998{\natexlab{a}})}]{dl98a}
{Draine}, B.~T. \& {Lazarian}, A. 1998{\natexlab{a}}, \apjl, 494, L19

\bibitem[{{Draine} \& {Lazarian}(1998{\natexlab{b}})}]{dl98b}
---. 1998{\natexlab{b}}, \apj, 508, 157

\bibitem[{{Draine} \& {Lazarian}(1999)}]{dl99}
---. 1999, \apj, 512, 740

\bibitem[{{Draine} \& {Lee}(1984)}]{drainelee84}
{Draine}, B.~T. \& {Lee}, H.~M. 1984, \apj, 285, 89, erratum: 1987 \apj, 318,
  485

\bibitem[{Draine \& Weingartner(1996)}]{draine96}
Draine, B.~T. \& Weingartner, J.~C. 1996, \apj, 470, 551

\bibitem[{Draine \& Weingartner(1997)}]{draine97}
---. 1997, \apj, 480, 633

\bibitem[{{Finkbeiner}(2004)}]{fink04}
{Finkbeiner}, D.~P. 2004, \apj, 614, 186

\bibitem[{{Finkbeiner} {\etal}(1999){Finkbeiner}, {Davis}, \&
  {Schlegel}}]{fds99}
{Finkbeiner}, D.~P., {Davis}, M., \& {Schlegel}, D.~J. 1999, \apj, 524, 867

\bibitem[{{Finkbeiner} {\etal}(2004){Finkbeiner}, {Langston}, \&
  {Minter}}]{flm04}
{Finkbeiner}, D.~P., {Langston}, G.~I., \& {Minter}, A.~H. 2004, \apj, 617, 350

\bibitem[{{Gold}(1952)}]{gold52}
{Gold}, T. 1952, \mnras, 112, 215

\bibitem[{{Goodman} {\etal}(1995){Goodman}, {Jones}, {Lada}, \&
  {Myers}}]{goodman95}
{Goodman}, A.~A., {Jones}, T.~J., {Lada}, E.~A., \& {Myers}, P.~C. 1995, \apj,
  448, 748

\bibitem[{Harper {\etal}(2000)Harper, Allen, Amato, Ames, Bartels, Casey,
  Derro, Evans, Gatley, Heimsath, Hermida, Jhabvala, Kastner, Loewenstein,
  Moseley, Pernic, Rennick, Rhody, Sandford, Shafer, Shirron, Voellmer, Wang,
  \& Wirth}]{hawc}
Harper, D.~A., \etal\ 2000, in \procspie\ 4014, Airborne
  Telescope Systems, ed. R.~K. Melugin \& H.-P. Roeser, 43

\bibitem[{{Heiles}(2000)}]{heilescat}
{Heiles}, C. 2000, \aj, 119, 923

\bibitem[{Hildebrand \& Kirby(2004)}]{aod}
Hildebrand, R. \& Kirby, L. 2004, in ASP Conf.\ Ser. 309, Astrophysics of Dust,
  ed. A.~N. Witt, G.~C. Clayton, \& B.~T. Draine, 515

\bibitem[{Hildebrand(1983)}]{rhh83}
Hildebrand, R.~H. 1983, \qjras, 24, 267

\bibitem[{Hildebrand(1988)}]{rhh88}
---. 1988, \qjras, 29, 327

\bibitem[{Hildebrand(2001)}]{tenerife}
Hildebrand, R.~H. 2001, in Astrophysical Spectropolarimetry, ed.
  J.~Trujillo-Bueno, F.~Moreno-Insertis, \& F.~Sanchez (Cambridge: Cambridge
  University Press), 265

\bibitem[{Hildebrand {\etal}(1999)Hildebrand, Dotson, Dowell, Schleuning, \&
  Vaillancourt}]{pspec}
Hildebrand, R.~H., Dotson, J.~L., Dowell, C.~D., Schleuning, D.~A., \&
  Vaillancourt, J.~E. 1999, \apj, 516, 834

\bibitem[{Hildebrand \& Dragovan(1995)}]{shape}
Hildebrand, R.~H. \& Dragovan, M. 1995, \apj, 450, 663

\bibitem[{{Hinshaw} {\etal}(2003){Hinshaw}, {Barnes}, {Bennett}, {Greason},
  {Halpern}, {Hill}, {Jarosik}, {Kogut}, {Limon}, {Meyer}, {Odegard}, {Page},
  {Spergel}, {Tucker}, {Weiland}, {Wollack}, \& {Wright}}]{hinshaw03}
{Hinshaw}, G., \etal\ 2003, \apjs, 148, 63

\bibitem[{{Houde} {\etal}(2004){Houde}, {Dowell}, {Hildebrand}, {Dotson},
  {Vaillancourt}, {Phillips}, {Peng}, \& {Bastien}}]{houde04}
{Houde}, M., {Dowell}, C.~D., {Hildebrand}, R.~H., {Dotson}, J.~L.,
  {Vaillancourt}, J.~E., {Phillips}, T.~G., {Peng}, R., \& {Bastien}, P. 2004,
  \apj, 604, 717

\bibitem[{{Kogut} {\etal}(1996){Kogut}, {Banday}, {Bennett}, {Gorski},
  {Hinshaw}, \& {Reach}}]{kogut96}
{Kogut}, A., {Banday}, A.~J., {Bennett}, C.~L., {Gorski}, K.~M., {Hinshaw}, G.,
  \& {Reach}, W.~T. 1996, \apj, 460, 1

\bibitem[{{Lazarian}(2003)}]{lazarian03}
{Lazarian}, A. 2003, Journal of Quantitative Spectroscopy and Radiative
  Transfer, 79, 881

\bibitem[{Lazarian \& Cho(2005)}]{lazcho_conf}
Lazarian, A. \& Cho, J. 2005, in ASP Conf.\ Ser.\ 343, Astronomical Polarimetry
  --- Current Status and Future Directions, ed. A.~Adamson, in press

\bibitem[{{Lazarian} \& {Draine}(1997)}]{ld97}
{Lazarian}, A. \& {Draine}, B.~T. 1997, \apj, 487, 248

\bibitem[{{Lazarian} \& {Draine}(1999{\natexlab{a}})}]{ld99a}
---. 1999{\natexlab{a}}, \apjl, 520, L67

\bibitem[{{Lazarian} \& {Draine}(1999{\natexlab{b}})}]{ld99b}
---. 1999{\natexlab{b}}, \apjl, 516, L37

\bibitem[{{Lazarian} \& {Draine}(2000)}]{ld00}
---. 2000, \apjl, 536, L15

\bibitem[{{Lazarian} \& {Finkbeiner}(2003)}]{lazfink}
{Lazarian}, A. \& {Finkbeiner}, D. 2003, New Astronomy Review, 47, 1107

\bibitem[{{Lazarian} {\etal}(1997){Lazarian}, {Goodman}, \& {Myers}}]{lgm97}
{Lazarian}, A., {Goodman}, A.~A., \& {Myers}, P.~C. 1997, \apj, 490, 273

\bibitem[{{Lazarian} \& {Yan}(2004)}]{lazarian04}
{Lazarian}, A. \& {Yan}, H. 2004, in ASP Conf. Ser. 309: Astrophysics of Dust,
  ed. A.~N. Witt, G.~C. Clayton, \& B.~T. Draine, 479

\bibitem[{{Lee} \& {Draine}(1985)}]{leedraine85}
{Lee}, H.~M. \& {Draine}, B.~T. 1985, \apj, 290, 211

\bibitem[{{Leitch} {\etal}(1997){Leitch}, {Readhead}, {Pearson}, \&
  {Myers}}]{leitch97}
{Leitch}, E.~M., {Readhead}, A.~C.~S., {Pearson}, T.~J., \& {Myers}, S.~T.
  1997, \apjl, 486, L23

\bibitem[{Li {\etal}(2006)Li, Novak, Chuss, Davidson, Dotson, Dowell,
  Hildebrand, Houde, Kirby, Krejny, Lazarian, Vaillancourt, \&
  Yusef-Zadeh}]{sharp2}
Li, H., \etal\ 2006, in Proc.\ SPIE, Astronomical
  Telescopes and Instrumentation, Orlando, FL, 24-31 May 2006, in prep.

\bibitem[{{Low} {\etal}(1984){Low}, {Young}, {Beintema}, {Gautier},
  {Beichman}, {Aumann}, {Gillett}, {Neugebauer}, {Boggess}, \&
  {Emerson}}]{low84}
{Low}, F.~J., \etal\ 1984, \apjl, 278, L19

\bibitem[{Martin(2006)}]{martin}
Martin, P.~G. 2006, this volume, astro-ph/0606430

\bibitem[{{Matthews} {\etal}(2003){Matthews}, {Chuss}, {Dotson}, {Dowell},
  {Hildebrand}, {Johnstone}, \& {Vaillancourt}}]{brenda-pspec}
{Matthews}, B.~C., {Chuss}, D., {Dotson}, J., {Dowell}, D., {Hildebrand}, R.,
  {Johnstone}, D., \& {Vaillancourt}, J. 2003, in Chemistry as a Diagnostic of
  Star Formation, ed. C.~L. Curry \& M.~Fich, 145

\bibitem[{{Matthews} {\etal}(2001){Matthews}, {Wilson}, \& {Fiege}}]{mwf01}
{Matthews}, B.~C., {Wilson}, C.~D., \& {Fiege}, J.~D. 2001, \apj, 562, 400

\bibitem[{Netterfield(2006)}]{netterfield}
Netterfield, B. 2006, this volume

\bibitem[{Novak {\etal}(2004)Novak, Chuss, Davidson, Dotson, Dowell,
  Hildebrand, Houde, Kirby, Krejny, Lazarian, Li, Moseley, Vaillancourt, \&
  Yusef-Zadeh}]{sharp}
Novak, G., \etal\ 2004, in Proc.\ SPIE
  5498, Millimeter and Submillimeter Detectors for Astronomy II, ed.
  J.~Zmuidzinas, W.~S. Holland, \& S.~Withington, 278

\bibitem[{Pollack {\etal}(1994)Pollack, Hollenbach, Beckwith, Simonelli,
  Roush, \& Fong}]{pollack94}
Pollack, J.~B., Hollenbach, D., Beckwith, S., Simonelli, D.~P., Roush, T., \&
  Fong, W. 1994, \apj, 421, 615

\bibitem[{Ponthieu(2006)}]{ponthieu}
Ponthieu, N. 2006, this volume

\bibitem[{Purcell(1979)}]{purcell79}
Purcell, E.~M. 1979, \apj, 231, 404

\bibitem[{{Rao} {\etal}(1998){Rao}, {Crutcher}, {Plambeck}, \&
  {Wright}}]{rao98}
{Rao}, R., {Crutcher}, R.~M., {Plambeck}, R.~L., \& {Wright}, M.~C.~H. 1998,
  \apjl, 502, L75

\bibitem[{Roberge(2004)}]{roberge04}
Roberge, W.~G. 2004, in ASP Conf.\ Ser. 309, Astrophysics of Dust, ed. A.~N.
  Witt, G.~C. Clayton, \& B.~T. Draine, 467

\bibitem[{Schleuning {\etal}(2000)Schleuning, Vaillancourt, Hildebrand,
  Dowell, Novak, Dotson, \& Davidson}]{w3}
Schleuning, D.~A., Vaillancourt, J.~E., Hildebrand, R.~H., Dowell, C.~D.,
  Novak, G., Dotson, J.~L., \& Davidson, J.~A. 2000, \apj, 535, 913

\bibitem[{{Tauber}(2004)}]{tauber04}
{Tauber}, J.~A. 2004, in The Magnetized Interstellar Medium, ed. B.~Uyaniker,
  W.~Reich, \& R.~Wielebinski, 191

\bibitem[{Vaillancourt(2002)}]{mythesis}
Vaillancourt, J.~E. 2002, \apjs, 142, 53

\bibitem[{{Watson} {\etal}(2005){Watson}, {Rebolo},
  {Rubi{\~n}o-Mart{\'{\i}}n}, {Hildebrandt}, {Guti{\'e}rrez},
  {Fern{\'a}ndez-Cerezo}, {Hoyland}, \& {Battistelli}}]{watson05}
{Watson}, R.~A., {Rebolo}, R., {Rubi{\~n}o-Mart{\'{\i}}n}, J.~A.,
  {Hildebrandt}, S., {Guti{\'e}rrez}, C.~M., {Fern{\'a}ndez-Cerezo}, S.,
  {Hoyland}, R.~J., \& {Battistelli}, E.~S. 2005, \apjl, 624, L89

\bibitem[{Whittet(2003)}]{whittetbook}
Whittet, D. C.~B. 2003, Dust in the Galactic Environment, 2nd edn., Series on
  Astronomy and Astrophysics (Philadelphia: Institute of Physics Publishing)

\bibitem[{Whittet(2004)}]{whittet04}
Whittet, D. C.~B. 2004, in ASP Conf.\ Ser. 309, Astrophysics of Dust, ed. A.~N.
  Witt, G.~C. Clayton, \& B.~T. Draine, 65

\bibitem[{{Whittet} {\etal}(2001){Whittet}, {Gerakines}, {Hough}, \&
  {Shenoy}}]{whittet01}
{Whittet}, D.~C.~B., {Gerakines}, P.~A., {Hough}, J.~H., \& {Shenoy}, S.~S.
  2001, \apj, 547, 872

\end{thebibliography}
\bibliographystyle{apj}


\end{document}